\documentclass[aps,prl,twocolumn,a4paper,10pt,notitlepage,footinbib,superscriptaddress,showpacs]{revtex4}%
\usepackage[english]{babel}
\usepackage[T1]{fontenc}
\usepackage{endnotes}
\usepackage{amssymb,amsmath,amsfonts}
\usepackage{textcomp}
\usepackage{graphicx,color}
\usepackage[dvipsnames]{xcolor}
\usepackage[utf8x]{inputenc}
\usepackage{bm}

\usepackage{comment}

\DeclareMathOperator*{\argmax}{argmax}

\newcommand\dmi[1]{{\color{black}#1}}
\newcommand\dmii[1]{{\color{black}#1}}

\begin{document}
\title{Dynamic and Facilitated Binding of Topoisomerase Accelerates Topological Relaxation} 

\author{D.~Michieletto}
\email{davide.michieletto@ed.ac.uk}
\affiliation{School of Physics and Astronomy, University of Edinburgh, Peter Guthrie Tait Road, Edinburgh, EH9 3FD, UK}
\affiliation{MRC Human Genetics Unit, Institute of Genetics and Cancer, University of Edinburgh, Edinburgh EH4 2XU, UK}

\author{Y.~G.~Fosado}
\email{yair.fosado@ed.ac.uk}

\author{E.~Melas}
\affiliation{School of Physics and Astronomy, University of Edinburgh, Peter Guthrie Tait Road, Edinburgh, EH9 3FD, UK}

\author{M.~Baiesi}
\affiliation{Department of Physics and Astronomy, University of Padova, 
Via Marzolo 8, I-35131 Padova, Italy}
\affiliation{INFN, Sezione di Padova, Via Marzolo 8, I-35131 Padova, Italy}

\author{L.~Tubiana}
\affiliation{Physics Department, University of Trento, via Sommarive, 14 I-38123 Trento, Italy}
\affiliation{INFN-TIFPA, Trento Institute for Fundamental Physics and Applications, I-38123 Trento, Italy}
\affiliation{Faculty  of  Physics,  University  of  Vienna,  Boltzmanngasse  5,  1090  Vienna,  Austria}

\author{E.~Orlandini}
\affiliation{Department of Physics and Astronomy, University of Padova, 
Via Marzolo 8, I-35131 Padova, Italy}
\affiliation{INFN, Sezione di Padova, Via Marzolo 8, I-35131 Padova, Italy}

\begin{abstract}
How type 2 Topoisomerase (TopoII) proteins relax and simplify the topology of DNA molecules is one of the most intriguing open questions in genome and DNA biophysics. Most of the existing models neglect the dynamics of TopoII which is characteristics for proteins searching their targets via facilitated diffusion. Here, we show that dynamic binding of TopoII speeds up the topological relaxation of knotted substrates by enhancing the search of the knotted arc. Intriguingly, this in turn implies that the timescale of topological relaxation is virtually independent of the substrate length. We then discover that considering binding biases due to facilitated diffusion on looped substrates steers the sampling of the topological space closer to the boundaries between different topoisomers yielding an optimally fast topological relaxation. We discuss our findings in the context of topological simplification in vitro and in vivo. 
\end{abstract}

\maketitle

\section{Introduction}

The topological regulation of DNA in vivo is a vital process that allows genes to be transcribed and cells to divide~\cite{Alberts2014}. The crucial role of Topoisomerases in mediating this regulation is now well established, and indeed they are among the most conserved proteins across living organisms~\cite{Wang2002}. Most surprisingly, type IIA Topoisomerase (TopoII) has been found to reduce the topological complexity of naked DNA below equilibrium in vitro~\cite{Rybenkov1997}. At the same time, recent works strongly suggest that yeast and even human DNA is itself surprisingly topologically simple in vivo~\cite{Valdes2018,Goundaroulis2019,Siebert2017}. While these fascinating findings have recently been partially explained in terms of a synergistic interaction of TopoII with Structural Maintenance of Chromosomes (SMC) proteins~\cite{Orlandini2019pnas,Racko2018,Dyson2020,Piskadlo2017a,Piskadlo2017}, a comprehensive model of TopoII-only topological regulation is still lacking.

Most of the existing \dmii{large-scale} polymer models of DNA and genome organisation mimic TopoII-bound regions by allowing inter-segment crossing at the expense of a modest amount of energy (comparable with the thermal one) and satisfy detailed balance~\cite{Goloborodko2016,Michieletto2014a,Flammini2004,Ziraldo2019,Witz2011,Levene2009}. 
A notable exception are the models accounting for hooked juxtapositions and/or violating detailed balance by restricting the directionality of strand-crossing~\cite{Huang2001,Vologodskii2001,Burnier2007,Liu2010} or by requiring a kinetic proofreading~\cite{Yan1999}. 
\dmii{However, none of these models consider the binding/unbinding dynamics of TopoII and how its potential bias in binding preference (e.g. to regions with large local DNA density) may affect the efficiency of TopoII to resolve DNA topology. To the best of our knowledge, addressing the question of how binding statistics and kinetics affect topological relaxation and topological steady states remains unexplored using large-scale polymer models}. Motivated by this, we specifically investigate the effect of dynamic binding of TopoII and its transitory binding biases due to, for instance, generic facilitated diffusion on locally looped substrates~\cite{Bressloff2013,Lomholt2009,Brackley2012,Brackley2013,Cagnetta2020}. 

The key discovery of this work is that both (un)binding kinetics 
and biased binding patterns to regions of large local DNA density dramatically shorten the timescale over which the topology of the substrates is relaxed to its steady state. Importantly, we find that the precise values of the (un)binding rates are not important as long as they are larger than the relaxation rate of the substrate. As we argue below, this may be thus a relevant regime in vivo, where the relaxation dynamics of the genome is dramatically slow~\cite{Rosa2008,Kang2015,Shi2018}.
Note that in this work we are not looking for a model of TopoII that can maintain the knotting probability below equilibrium at steady state, but rather focus on the effects of TopoII dynamic and facilitated (un)binding to/from the substrate. To the best of our knowledge this aspect has never been systematically and quantitatively investigated in either simulations or experiments.

\enlargethispage{-65.1pt}

By using a simplified 2D random walk model on idealised knot spaces, we argue that dynamic and biased binding enhance the sampling between knot types thereby speeding up the unknotting process. Strikingly, due to the (sublinear) scaling behaviour of the knotted region with polymer length~\cite{Marcone2005,Tubiana2011prl}, we also find that these binding mechanisms naturally yield a rate of topological relaxation that is virtually independent on the substrate length. Pleasingly, this is in line with recent experiments showing that the topological complexity of DNA in vivo is broadly independent on its length~\cite{Valdes2018}.

\section{MATERIALS AND METHODS}
we model a torsionally relaxed (nicked) DNA plasmid about $L=3.6$ kbp-long as a bead-spring polymer made of $N=500$ beads (each bead having size $\sigma=2.5 $nm $=7.3$ bp) connected in a ring. Inter-bead interactions are modelled with a purely steric truncated and shifted Lennard-Jones (LJ) repulsion 
\begin{equation}
    U_{LJ}(r) = 4 \epsilon \left[ (\sigma/r)^{12} - (\sigma/r)^6 \right] + \epsilon
\end{equation}
for $r<r_c=2^{1/6}\sigma$ and 0 otherwise. The TopoII-bound segment is modelled by allowing a 50 beads ($L=360$ bp)-long segment to undergo strand-crossing with a small energy penalty ($A=2 k_B T$). This is done by modelling the interactions between this and all the other beads with a soft potential 
\begin{equation}
U_s(r) = A (1 + \cos{\left( \pi r/r_c \right)} )
\end{equation}
for $r<r_c=2^{1/6}\sigma$ and 0 otherwise. \dmi{Note that to overcome this soft potential a bead must pay an energy penalty; this means that chain-crossing events occur according to a Boltzmann probability}. \dmi{Importantly, only the TopoII-bound segment displays soft interactions with the other beads, whereas the rest of the chain preserves fully repulsive LJ interactions thus preventing strand-crossing events that involve more than 2 strands.} Each bead is connected to its two neighbors along the ring by using a FENE potential
\begin{equation}
    U_{FENE}(r) = - 0.5 K R_0^2 \log{\left[1 - \left( r/R_0 \right)^2\right]} 
\end{equation}
with $K=30 \epsilon/\sigma^2$ and $R_0=1.6\sigma$. Finally, the semiflexible nature of DNA is modelled via a Kratky-Porod potential
\begin{equation}
    U_{bend}(r) = \dfrac{k_BT l_p}{\sigma} (1 + \cos{\theta})
\end{equation}
where $\theta$ is the angle defined by two consecutive bond vectors along the polymer and $l_p=20\sigma= 50$ nm is the persistence length~\cite{Calladine1997}. 

We have chosen the length of the TopoII-bound segment to be $50$ beads $\simeq 125$ nm (or about 2 persistence lengths) for computational efficiency as shorter segments (for instance 10 beads $\simeq 25$ nm) yield qualitative similar results but display slower kinetics and hence require longer simulations. We also mention that we have performed simulations in which the TopoII region would remain strongly bent during the topological simplification time and we could not see any appreciable difference with the non-bent case (see SI Fig.~S6). The dynamical update of the ``soft'' segment is done by updating the position of the TopoII in one of the following ways: either by a random jump, by diffusion, by a jump to the segment with maximum local curvature or by a jump to the segment of maximum local density. We have also compared these dynamic biased binding models against their static versions (see SI Fig.~S5).\dmii{We note that to avoid numerical instabilities we model TopoII unbinding by first increasing the soft energy barrier to $A = 20 k_BT$ for a short time and then by switching the segment back to fully repulsive LJ potential.}

For these different models of TopoII binding dynamics we study the relaxation of the knotting probability $P_K(t)$ to its equilibrium value starting from a DNA molecule pre-knotted into a $5_1$ torus knot (unless otherwise stated). The averages of the relaxation process are performed over at least 64, and up to 400, independent replicas. We follow the topology of the polymer by computing its Alexander determinant using the kymoknot software~\cite{Tubiana2011,Tubiana2018} and compute the knotting probability $P_K(t)$ by counting the number of replicas that are knotted at a given time. \dmi{We recall that we work always with only one TopoII bound on the ring at any one time, so the knotting relaxation curves should be considered a lower bound in the case the protein was present in stoichiometric excess. On the other hand, it is always possible in an in vitro experiment, to tune the stoichiometry so that only one TopoII (on average) is bound to each plasmid at any one time}. Since the knotting probability is computed over a binary value (either knotted or unknotted), the errors on the mean are obtained by the blocking method~\cite{Newman1999}, i.e.~by randomly assigning simulation replicas to different partitions, computing the mean knotting probability for each partition and finally computing the standard error of the mean (SEM) across partitions.

The molecular dynamics with implicit solvent (Langevin) simulations are evolved within the LAMMPS~\cite{Plimpton1995} engine coupled to custom-made C\texttt{++} codes to perform the dynamic update of the TopoII region. (We provide sample codes in a GitHub repository, see below). Note that we pre-equilibrate the chains at fixed topology (i.e.~without TopoII) for at least one relaxation time of the chain, i.e. $10^5$ $\tau_B$, where $\tau_B$ is the Brownian time (see SI, Figs.~S1-S5).   

To map our simulation timescales to real time we consider the Brownian time of a bead, i.e. $\tau_B=\sigma^2/D = 3\pi \eta_w\sigma^3/k_BT = 40$ ns (assuming $\eta_w=1$ cP as the viscosity of water) finding that the chain relaxation time is about $\tau_R \simeq 10^5 \tau_B \simeq 4$ ms.  We can also compare our effective dissociation constant with that found in the literature for TopoII as follows. Consider an off-rate $k_{\rm off} = 10^{-5} \tau_B^{-1} = 250 s^{-1}$ and an estimate for the diffusion-controlled on-rate $k_{on} \simeq 4 \pi (D_1 + D_2) (a_1 + a_2) \simeq 5 \times 10^9 M^{-1}  s^{-1}$ with $D_1 = k_BT/(3 \pi \eta_w a_1)\simeq 10 \mu m^2/s$, $D_2 \ll D_1$, $a_1=40$ nm the size of TopoII~\cite{Hizume2007,Schultz1996} and $a_2=10$ nm the size of the target DNA site. These yield an equilibrium dissociation constant $K_d = k_{\rm off}/k_{on} \simeq 50$ nM which is in line with the value found for TopoII on relaxed DNA (36 nM)~\cite{Charvin2005a}. 

We note that papers in the literature attempting to measure the residence time of TopoII in vivo using fluorescence recovery after photobleaching (FRAP) report values in the 1-20 seconds range, or $k_{off} \simeq 0.1-1 s^{-1}$~\cite{Kalfalah2011,Christensen2002}. \dmii{In vivo, we can take this into account using the nucleoplasm viscosity $\eta_n= 1000$ cP (1000$\times$ water viscosity~\cite{Caragine2018}) which gives $k_{on} \simeq 2 \times 10^7 M^{-1} s^{-1}$; in turn we obtain $K_d = 5-50$ nM which is again in line with typical dissociation constant for TopoII~\cite{Charvin2005a}.} In spite of this, FRAP results are known to vary widely depending, among other things, on the size of the bleached spot and the kinetic model used to fit the FRAP curves~\cite{Mazza2012}. More precise measurements on the residency time of TopoII on DNA would be highly welcomed. 

We highlight that even if the dissociation constant were smaller, e.g. $K_d \simeq 0.1$ nM due to small unbinding rates $k_{\rm off} = \tau_j^{-1} \simeq 0.1- 1$ $s^{-1}$ {\it in vitro} (water viscosity) the fact that the relaxation time of a DNA molecule grows as $\sim N^2$ in dilute conditions and as $\sim N^3$ in dense conditions~\cite{Doi1988}, ensures that there must be regimes in which $\tau_j \leq \tau_R$ for sufficiently long DNA molecules. Even more interestingly, it is well accepted that the relaxation time of genomes may be on the order of minutes or hours~\cite{Rosa2008,Shi2018,Kang2015}. All this renders our findings for $\tau_j \leq \tau_R$ relevant for the topological relaxation of large DNA molecules both \emph{in vitro} and \emph{in vivo}.  


\section{Results}

\begin{figure*}[t!]
\centering
\includegraphics[width=0.8\textwidth]{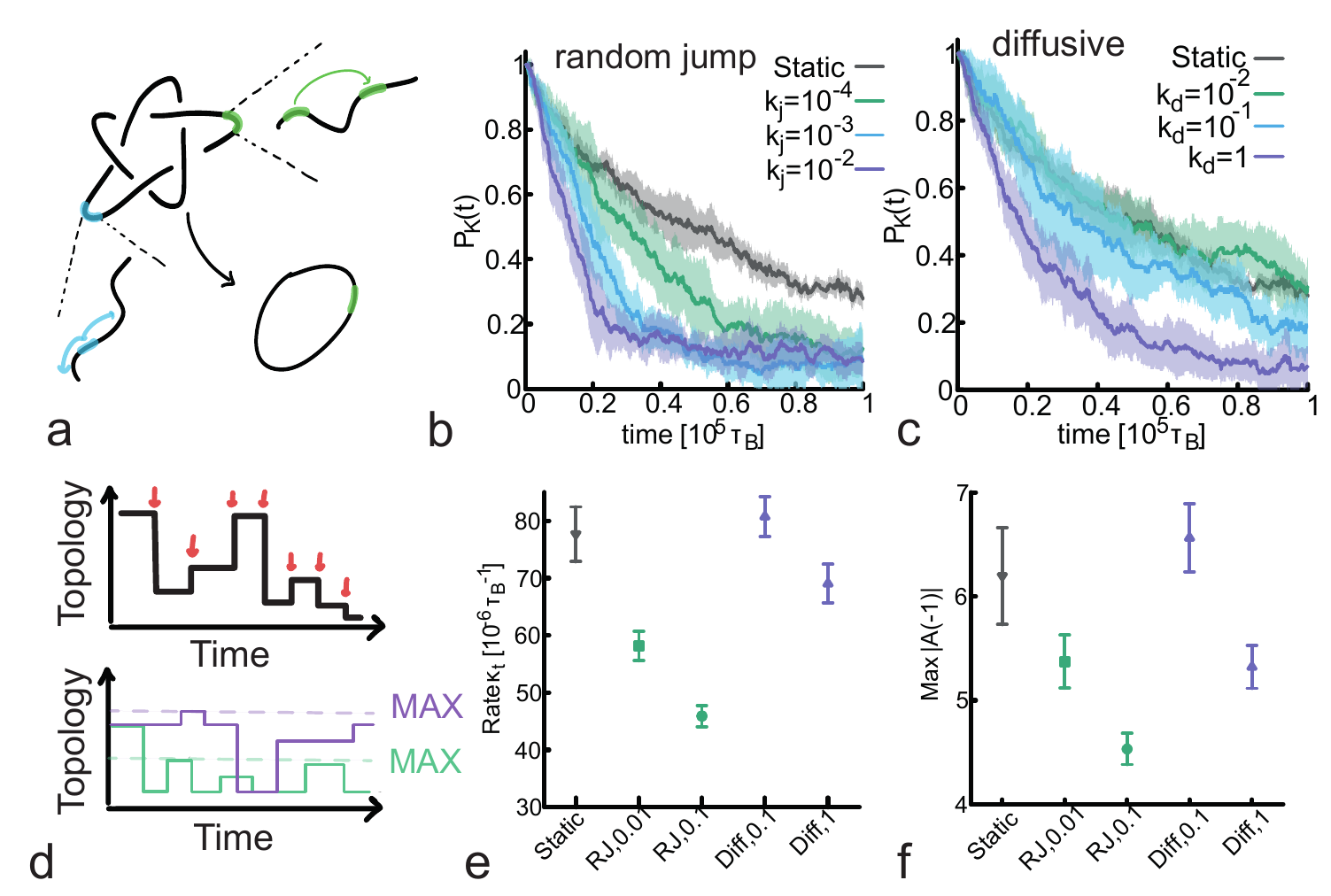}
\vspace{-0.2 cm}
\caption{\textbf{Unknotting is faster for dynamic TopoII.} \textbf{a} Sketch of the system: one TopoII remains bound to the DNA and changes location by either random jumps (green) at rate $k_j$ (units of $\tau_{B}^{-1}$) or via diffusion (orange) at rate $k_d$ (units of $\sigma^2/\tau_{B}$). \textbf{b} The larger the jumping rate of TopoII, the faster the relaxation of $P_K(t)$. \textbf{c} The relaxation of $P_K(t)$ is also faster for more diffusive TopoII, albeit the effect is milder than in \textbf{b}. \textbf{d} Sketch of how we compute $\kappa_t$ and $\textrm{Max} |A(-1)|$. \textbf{e} $\kappa_t$ is model and rate dependent and decreases for larger jumping/diffusion rates. \textbf{f} $\textrm{Max} |A(-1)|$ also decreases for more dynamic binding or faster diffusion. Smaller $\kappa_t$ and $\textrm{Max} |A(-1)|$ indicate slower but more precise evolution of topology towards the steady state, overall rendering the simplification process more efficient.
}
\label{fig:dynamic1}
\vspace{-0.2 cm}
\end{figure*}


\subsection{Random Dynamic Binding}
To model the (un)binding kinetics of TopoII (from) to a DNA substrate, we consider a simplified scenario in which there is always one TopoII bound, and analyse the relaxation of the knotting probability for different rates at which TopoII is relocated. 

For a random (un)binding process, the constraint that there is only one TopoII bound is equivalent to the situation in which TopoII randomly ``jumps'' at rate $k_j$ (with $k_j = 0$ being the static case) measured in (inverse) Brownian times $\tau_B=\sigma^2/D_0$ (see Materials and Methods). From our simulations, a rate as little as $k_j=10^{-4} \tau_B^{-1}$ already appears to give a significant deviation from the static case. Rates equal, or larger than, $k_j=10^{-2} \tau_B^{-1}$  have reached the maximum relaxation efficiency (see Fig.~\ref{fig:dynamic1}b). Since the longest relaxation time of the chain under consideration ($N=500$ beads and $l_p=20\sigma$ in this figure) is about $10^5 \tau_B$ (see Materials and Methods and SI), it means that about 10 ``jumps'' within one relaxation time are enough to significantly speed up the topological simplification with respect to a static TopoII. 

For instance, for a $N \simeq 300$ kbp-long DNA molecule in free solution with $R_g \simeq 2$ $\mu$m~\cite{Robertson2006} and diffusion coefficient $D \simeq 0.1$ $\mu$m$^2$/s~\cite{Robertson2006} one finds a relaxation time $\tau_R = R_g^2/D \simeq 20$ seconds. This implies that a residency time of 2 seconds, or a $k_{\rm off} \simeq 0.5$ s$^{-1}$ ($K_d \simeq 0.1$ nM) would display a strong enhancement of the topological simplification compared with a perfectly immobile TopoII. This may be tested experimentally as residency time of TopoII on DNA is expected to be salt dependent (see also Conclusions section). 

Motivated by the notion of facilitated diffusion, namely a target search made by alternated rounds of 3D and 1D diffusion~\cite{Brackley2012,Lomholt2009,Loverdo2009,Mirny2009a}, we also test whether and to what extent a purely curvilinear (1D) diffusion of TopoII impacts on the topological simplification rate. Similarly to the case of random jumps, we update the position of the only TopoII bound by sliding it randomly along the substrate every $k_d^{-1}$ time step, i.e. with a diffusion constant $D = 0.5 k_d \sigma^2$. We find that larger diffusion rates ($k_d \geq \tau_B^{-1})$ are needed to relax as efficiently as random jumps (Fig.~\ref{fig:dynamic1}c). This entails that facilitated diffusion combining 3D jumps and 1D sliding~\cite{Mirny2009a} may be the best strategy to simplify the topology of the substrate.

\dmi{To better understand and quantify the observed enhancement in topological simplification, we compute two topological observables: (i) the rate (number of events per unit time) $\kappa_t$ at which topology-changing strand-crossing operations occur, without discriminating between events that reduce or increase the knot complexity and (ii) the maximum value attained by the Alexander determinant $|A(x)|$ evaluated at -1 (a topological invariant of the system and also known as knot determinant~\cite{michieletto2020separation}) during the course of each simulation, denoted as $\textrm{Max} |A(-1)|$. We note that up to 6-crossing knots, its value is directly correlated to the knot complexity~\cite{DAdamo2015a,Dabrowski-Tumanski2021}. These observables are schematically shown in Fig.~\ref{fig:dynamic1}d. Practically, the rate $\kappa_t$ is measured by counting the number of times we observe a change in topology during the course of a simulation (indicated schematically by red arrows in Fig.~\ref{fig:dynamic1}d) and by dividing that number by the total simulation time. $\textrm{Max} |A(-1)|$ is instead computed by taking the maximum of the Alexander determinant evaluated at -1 over the course of each simulation and excluding the initial $5_1$ topology (skematically shown by the ``MAX'' horizontal dashed lines in Fig.~\ref{fig:dynamic1}d). These two topological observables yield complementary information. The former, $\kappa_t$, is a measure of how often the system attempts to simplify its topology while the latter, $\textrm{Max} |A(-1)|$, is a measure of how widely the substrate explores the attainable topological space.}

\dmi{In Fig.~\ref{fig:dynamic1}e one can appreciate that both the diffusive and jump models yield a smaller $\kappa_t$ compared to the static one. Thus, interestingly, dynamic binding of TopoII gives rise to processes in which the knot type of the substrate changes less frequently than in the case of static TopoII. We explain this by noting that a static TopoII allows for rapid and repeated strand-crossing events which increase the rate $\kappa_t$. On the other hand, faster jumping rate reduce the probability of repeated strand-crossing events, hence $\kappa_t$ becomes smaller. 
At the same time, we also report that $\textrm{Max} |A(-1)|$ depends on the underlying binding process and the corresponding rates (Fig.~\ref{fig:dynamic1}f). In particular, it is smaller for fast diffusion and for moderately fast (compared with the Brownian time) random jumps. This observation is less straightforward to explain. We argue that this is due to the fact that binding of TopoII effectively creates a locally denser region of the substrate because the bound region starts to behave as an ideal phantom chain rather than a self-avoiding walk. This change in local statistics of the chain affects the knotting probability and increases the chances of creating complex topologies. On the contrary, in the case in which TopoII frequently changes its position along the chain, there is no time for the chain to locally crumpled and change its statistics to that of a phantom chain and hence prevents the formation of complex knots, a behaviour mirrored by the lower values displayed by $\textrm{Max} |A(-1)|$. }

\begin{figure*}[t!]
\centering
\includegraphics[width=0.7\textwidth]{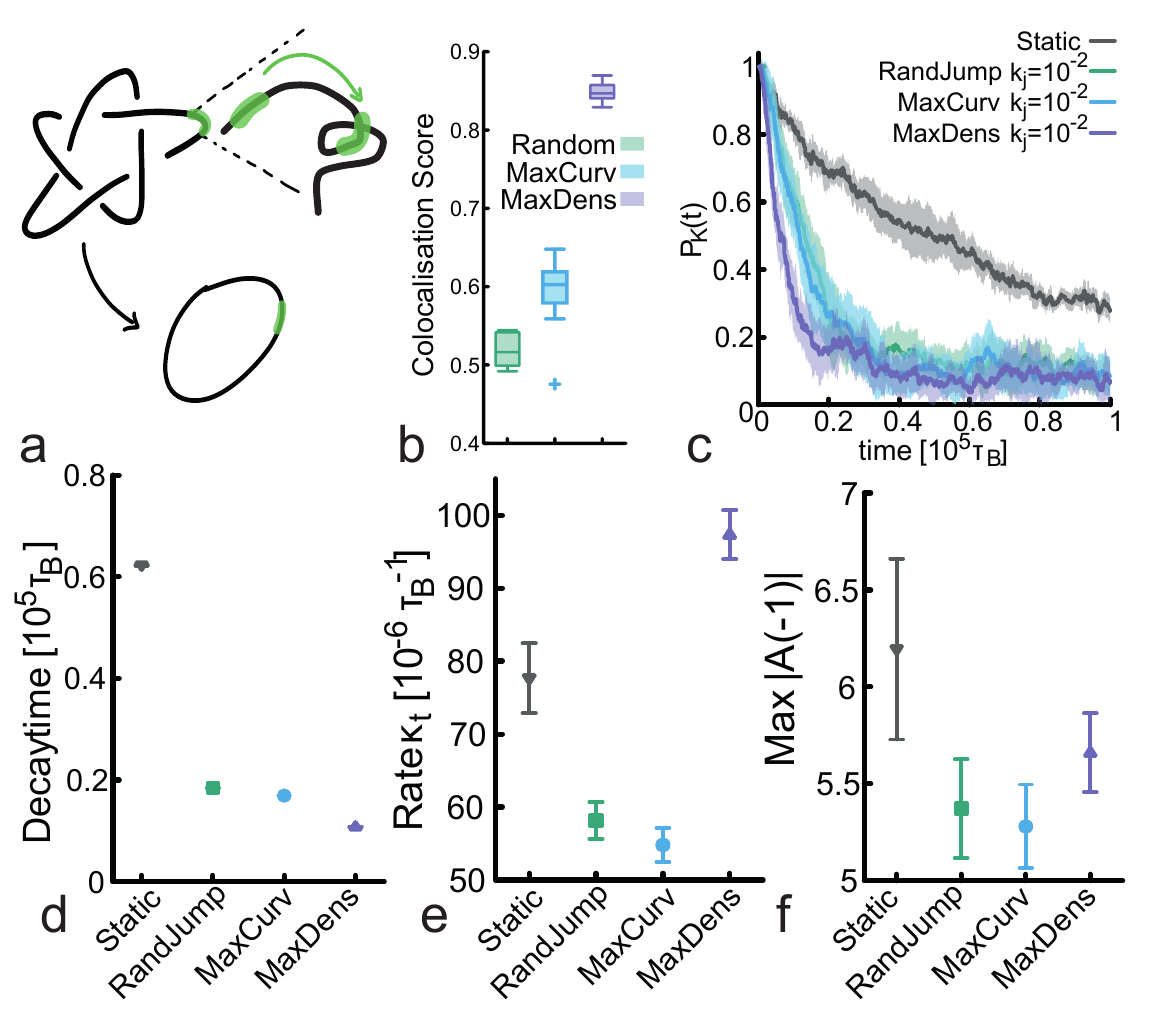}
\vspace{-0.2 cm}
\caption{\textbf{TopoII binding to Regions of Local Maximum Density is the Most Efficient Simplification Pathway.} \textbf{a} Sketch of the model. \textbf{b} Boxplot showing that, on a $5_1$ knot spanning about half of the contour length of polymer, the bead with maximum local density is more than 80\% of the times colocalised with the knotted arc, in agreement with the weakly localised nature of the knot. \textbf{c} Knotting probability for different models and same jumping rate. \textbf{d} Decay time $\tau$ obtained from an exponential fitting $e^{-t/\tau}$ of the knotting probability curves in \textbf{c} at early times. \textbf{e} Rate of topology-changing operations. \textbf{f} Average maximum value of Alexander determinant at -1.  }
\label{fig:compare_dynamic_maxdens}
\vspace{-0.2cm}
\end{figure*}

In summary, the behavior of $\kappa_t$ and $\textrm{Max} |A(-1)|$ suggest that dynamic binding of TopoII yields more precise relaxation pathways involving fewer topology-changing strand-crossings. This may be explained as follows: once TopoII lands on a segment embedded within a physically knotted region (or knotted arc)~\cite{Marcone2005,Tubiana2011prl}, there is an entropic gain in relaxing the substrate to its topological equilibrium. On the other hand, if TopoII lands on a region that is not physically knotted and sufficiently swollen, then the chances of generating a topology-changing strand crossing in a short time (before TopoII moves somewhere else) is low. Furthermore, the limited residence time hinders repeated topology-changing strand-crossing events. 

Finally, we should stress that in static models of TopoII, the rate-limiting step for the topological evolution of the substrate is the diffusion dynamics of the knotted region (i.e.~the knot has to find the TopoII-bound segment). In analogy with reptation polymer dynamics, this process scales as $\sim (N-l_k)^2/D_k$ with $l_k$ being the length of the knot and $D_k\simeq D_0/l_k$ its curvilinear diffusion~\cite{Bao2003,Doi1988}. On the other hand, for dynamic TopoII, it is the protein that ``finds'' the knot first, in a much shorter timescale $\lesssim N/k_j$.

\subsection{Biased and Facilitated Binding} 
Motivated by the fact that TopoII has been proposed to co-localise with DNA regions of large curvature~\cite{Hardin2011,Vologodskii2001,Burnier2007}, we explore a model in which the protein binds preferentially to regions with high bending. Additionally, we also investigate the case in which TopoII preferentially binds regions of large local density. These two binding modes are motivated by the fact that, in general, any protein that displays a weak non-specific binding affinity to DNA is expected to undergo facilitated diffusion~\cite{Mirny2009a} which can create transient binding biases to regions of crumpled or looped DNA~\cite{Lomholt2009} (i.e. with large density) or, as we argue here, knotted DNA due to its (weakly) localised~\cite{Tubiana2013}, or even metastable tight~\cite{Grosberg2007}, nature. 

To couple the binding dynamic process with these geometric features (i.e.~curvature and local density) we define the region of maximum local curvature as the portion of the polymer around the bead of index $i_{maxcurv}$ such that 
\begin{equation}
i_{mc} = \argmax_i \left[ \sum_{j=-l_T/2}^{l_T/2} 1 - \dfrac{\bm{t}_{i+j} \bm{t}_{i+j+1}}{|\bm{t}_{i+j}| |\bm{t}_{i+j+1}|} \right]
\end{equation} 
where $l_T=50$ is the length of the TopoII region, $\bm{t}_i \equiv \bm{r}_{i+1}-\bm{r}_{i}$ is the tangent vector at bead $i$ (periodic conditions on the index $j$ are implicit). Similarly, the region of maximum local density is defined as 
\begin{equation}
i_{md} = \argmax_i \left\{ \sum_{j=1}^N \Theta(R - |\bm{r}_i - \bm{r}_j|)  \right\}
\end{equation} 
where $\Theta(x)=1$ if $x>0$ and 0 otherwise and we take $R=50^\nu$ with $\nu=0.588$ as expected for a chain with excluded volume interaction~\cite{Doi1988}. Interestingly we find that for a $5_1$ topology spanning about 50\% of the polymer ($N=500$, $l_p = 20\sigma$), $i_{mc}$ and $i_{md}$ are located within the physically knotted region about 60\% and 85\% of the times, respectively (Fig.~\ref{fig:compare_dynamic_maxdens}b). We highlight that both these biases violate detailed balance because they systematically move TopoII to $i_{mc}$ and $i_{md}$ without any intermediate steps. These models can thus be interpreted as extreme realisations of thermodynamic biases due to, for instance, conformation-driven docking~\cite{Hardin2011} and transient enrichment in locally folded regions due to facilitated diffusion~\cite{Lomholt2009,Mirny2009a}. Indeed, while on ideal uniform substrates the binding rate is given by the Smoluchowski constant $k_{on} = 4 \pi (D_1 + D_2) (a_1 + a_2)$, within a coiled substrate this binding rate depends on the typical distance between neighbouring segments $l \simeq \rho^{1/3}$. Thus, on timescales shorter than the chain relaxation time, a TopoII with non-specific attraction to DNA will undergo a significant oversampling of large density regions before exploring the rest of the substrate~\cite{Lomholt2009}. Even more intriguingly, simulations show that permanently looped or crumpled regions may act as sinks for weakly sticky proteins searching their target~\cite{Brackley2012}. \dmi{To link this to the specific case of TopoII, we highlight that TopoII has two DNA-binding sites (associated to the ``gate'' and ``transfer'' DNA segments~\cite{Bates2005}) and will thus minimise its free energy if both are bound to DNA. This can be most easily achieved in regions with increased local concentration of DNA, which offer a larger amount of binding site per unit volume.}

In Fig.~\ref{fig:compare_dynamic_maxdens}c we compare the relaxation of the knotting probability for these different mechanisms: \dmi{random} static, random jumps or jumps biased to regions of maximum curvature or maximum density (with same jumping rate). Intriguingly, the fastest relaxation is achieved when TopoII preferentially binds to regions of maximum density (see Fig.~\ref{fig:compare_dynamic_maxdens}d for the decay time). \dmi{We also note that we have checked the case in which we place a static TopoII in regions of maximum density or curvature at time 0 and let the simulation evolve. These yield the same result as static TopoII randomly placed along the contour (see SI Fig.~S5). To exclude that the observed difference in relaxation rate depends on the size of the polymer, we additionally checked that the steady state radius of gyration of the ring is the same in the different models (see SI Fig.~S7).}

Measuring $\kappa_t$ and $\textrm{Max}|A(-1)|$ confirms that biases towards regions of maximum curvature enhance the simplification efficiency by reducing the number of topology-changing operations while rendering them more accurate, i.e.~both $\kappa_t$ and $\textrm{Max}|A(-1)|$ decrease. On the other hand, dynamic binding to regions of maximum density yields a qualitatively different relaxation pathway. As shown in Fig.~\ref{fig:compare_dynamic_maxdens}c-d, we find that, although the number of topology-changing operations per unit time $\kappa_t$ is much larger than for any of the other models, this feature is only accompanied by a mild increase in $\textrm{Max}|A(-1)|$ with respect to the others. In other words, binding to regions of high local density facilitates frequent strand-crossing operations but this is not accompanied by a significant enhancement of the knot complexity; on the contrary, it appears that most of these operations are still simplifying the topology.

\begin{figure}[t!]
\centering
\includegraphics[width=\columnwidth]{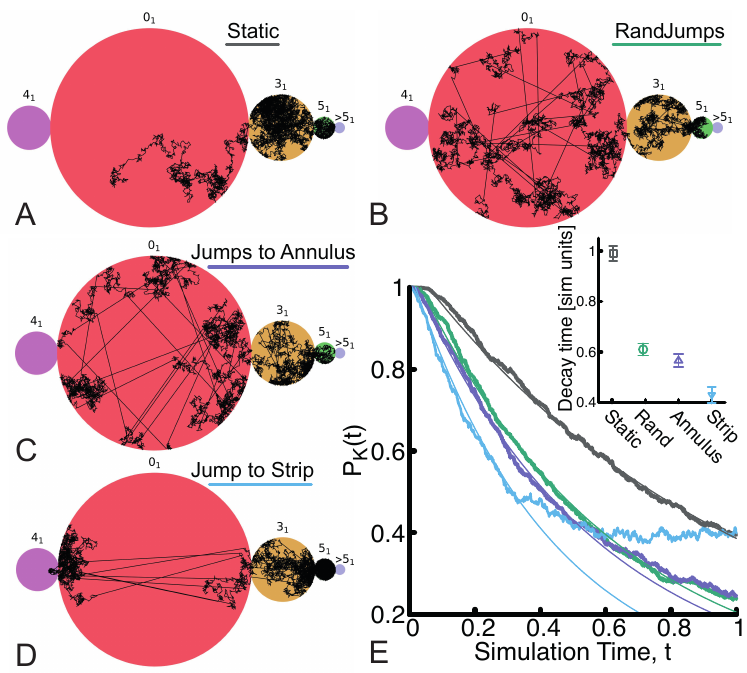}
\vspace{-0.3 cm}
\caption{\textbf{A simplified model of topological sampling}. \textbf{a-d} Example trajectories of walks within simplified topological spaces: (a) purely diffusive and (b-d) with jumps either random or closer to the annulus or the vertical strip near the boundary of the disk. All simulations are initialised from within the $5_1$ space (green disk). \textbf{e} Shows $P_K(t)$ curves generated by this simplified model which well capture the faster relaxation for walks with biased jumping. Solid lines show the fits an exponential function $e^{-t/\tau}$ at early times. The inset shows the decay time $\tau$.
}
\label{fig:model_sampling}
\vspace{-0.3 cm}
\end{figure}

\subsection{A simplified Random Walk kinetic model of DNA unknotting}

To understand qualitatively how the dynamics of TopoII affect its ability in exploring the possible topological spaces (set of configurations with the same knot type) we present a simple model in which each DNA configuration is described by its $3N$ coordinates and the index of the monomer where the TopoII is bound. When TopoII is bound we expect that the $3N+1$-dimensional space of configurations is locally explored by geometrical changes of the coordinates, i.e. by thermal motion of the polymer segments. On the other hand, a spatial relocation of TopoII would correspond to  a significant ``jump'' in the configurational space, even if the knot type is preserved, i.e. it would correspond to a jump within the same topological space. These features can be described by an idealised two-dimensional stochastic walk that explores a set of disks representing different knot spaces, see Fig.~\ref{fig:model_sampling} and Movies in SI. Each point within a disk represents one of the $3N+1$-dimensional configurations with a given knot type. Those bearing a TopoII in a position compatible with a topological change from knot $\mathcal{k}_1$ to knot $\mathcal{k}_2$ are mapped to points close to the boundary between disks representing the two knot types. Note that the size of each disk pictorially represents the number of possible polymer configurations with that knot type. A larger size corresponds to a knot type with larger entropy and hence more available configurations. We also note that while the dimensionality of the walk is important to determine its statistics, below we shall be interested in the kinetics of the sampling of these spaces. For instance, irrespectively of dimensionality $d$, a random walk with diffusion coefficient $D$ displays a mean squared displacement $\Delta r^2 (n) = 2d Dn$, with the $n$ the number of steps. This means that (at fixed dimensionality $d$) faster walks explore the available space more quickly, irrespectively of the dimension of the space. 

With these assumptions, we perform stochastic simulations in which 1000 random walks are initialised at the center of the $5_1$ disk, i.e. from the same polymer configuration with knot type $5_1$ that is far from being converted to a different knot type from a simple strand passage. The walks are allowed to diffuse following  a stochastic diffusive motion within the set of slightly overlapping disks.
This purely diffusional motion represents the DNA fluctuating in space with TopoII statically bound at a given site and is termed model (i). The motion of TopoII at fixed DNA configuration is modeled instead by a random jump of the walker within the current disk thus preserving the knot type. These jumps take place with constant rate so that on average about $300$ jumps are performed in a time unit. Model (ii) adds this feature to the normal diffusion of model (i). Thus, we qualitatively associate model (i) with static TopoII and model (ii) with random TopoII jumps.
At sufficiently long times they converge to the same equilibrium statistics because they are both reversible processes, yet kinetics (ii) is expected to be faster than (i) since it effectively displays a larger diffusion coefficient. Hence, model (ii) would be faster at finding the boundaries of the current disk (i.e. topological space), thus yielding a faster topological relaxation process. The numerical simulations of model (i) and model (ii) are indeed compatible with this observation (see Fig.~\ref{fig:model_sampling}e and compare grey with green curve). According to this picture TopoII proteins performing random jumps should be found more frequently at the boundaries between knot spaces and hence have more chances to simplify DNA topology. Clearly, and in agreement with the arguments we presented above, if the time between two random jumps is comparable to the time required by the walker to explore the full disk by simple diffusion (conformational relaxation time at fixed knot type) the topological simplification rate in presence of random jumps would be the same than that of its static counterpart.

To mimic biases toward regions of maximum curvature/density of the substrate, we make the assumption that points near the boundaries of the disks represent polymer configurations in which TopoII is bound in such a way that a short diffusion of the polymer segments can lead to a change in topology. This is implemented in the 2D walker model by forcing jumps to land within a thin annulus near the edge of the disk (model (iii)) and within a thin vertical strip closest to the boundary of contiguous disks (model (iv)) (see Fig.~\ref{fig:model_sampling}c,d respectively and corresponding the SI Movie). These biased jumps reflect our working hypothesis, i.e. that the regions closest to neighbouring knot spaces are populated by conformations that are more likely to precede a change in topology; among these, we argue there should be some in which TopoII is bound at regions of large local curvature/density. Thus, jumping to the annulus of a knot space or the vertical strip near the neighbouring space enhances the chances for the walker to make a topological transition. Model (iii) and (iv) follow the same principles of the random jumps model (ii) as they effectively give rise to an enhanced diffusion coefficient. In contrast to model (ii) though, they display an out-of-equilibrium irreversible dynamics as their jumping rules explicitly violate detailed balance. In this respect it is not surprising that both the kinetics and the long-time steady state of the knotting probability of model (iii) and (iv) are different from those displayed by the equilibrium models. In particular, we find that both (iii) and (iv) models yield faster topological relaxation kinetics (see purple and cyan curves in Fig.~\ref{fig:model_sampling}E) with model (iv) being the fastest at fixed jumping rate. 

To conclude, we find intriguing that a simplified model of a random walk in abstract two-dimensional knot spaces qualitatively capture the salient points that we observed in the more complicated MD simulations (compare Figs.~\ref{fig:dynamic1}b,c and ~\ref{fig:compare_dynamic_maxdens}c). This simplified models provides an intuitive and physically appealing interpretation, within the topological space representation, of the impact that different dynamics of TopoII on DNA substrates may have on their topological relaxation.
\color{black}

\subsection{Dependence of Topological Simplification Rate on Substrate Length}
\begin{figure}[t!]
\centering
\includegraphics[width=\columnwidth]{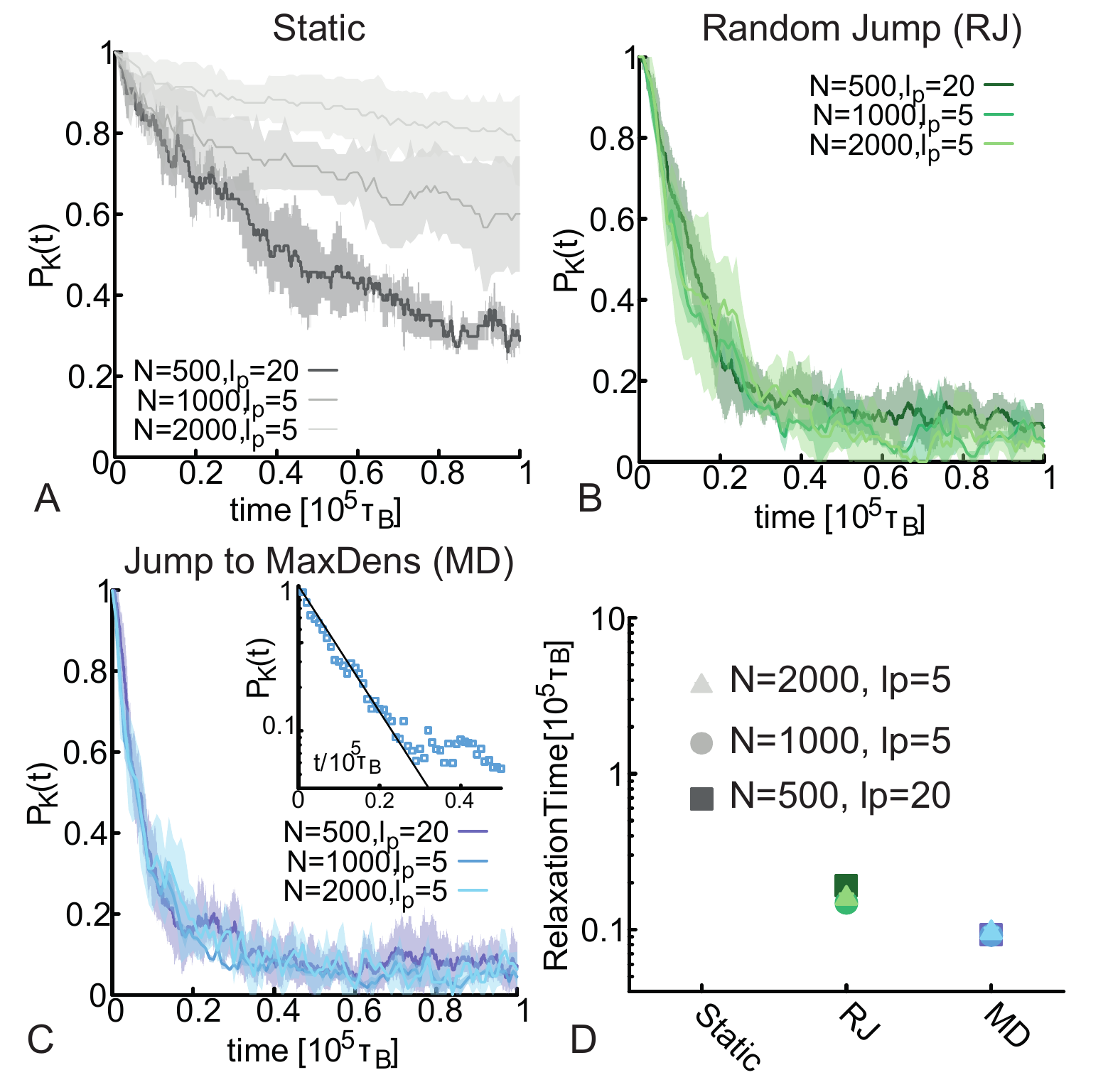}
\vspace{-0.7 cm}
\caption{\textbf{Dynamic Simplification Rate is independent on DNA Length}. \textbf{a} Simplification curves for Static TopoII. \textbf{b} Simplification curves for dynamic random jump, rate $k_j=10^{-2} \tau_B^{-1}$. \textbf{c} Simplification curves for dynamic jump to max density, rate $k_j=10^{-2} \tau_B^{-1}$. Inset shows the exponential fit $e^{-t/\tau}$ of one of the curves at short times. \textbf{d} Topological relaxation timescale plotted for the different systems. }
\label{fig:compare_lengths}
\vspace{-0.6 cm}
\end{figure}

Finally, we investigate the sensitivity of dynamic models against substrate length $N$ and flexibility. In Fig.~\ref{fig:compare_lengths} we show that, unlike the static model of TopoII, all the dynamic models yield relaxation curves that are very weakly dependent of $N$ and persistence length. Additionally, from Fig.~\ref{fig:compare_lengths}d it is apparent that biasing the binding dynamics towards region of maximum local density provides the most efficient way of performing topological relaxation, especially for very long substrates. 

We explain this weak dependence on substrate length (reminiscent of some recent experiments in vivo finding plateauing knot complexity at large DNA lengths~\cite{Valdes2018}) as follows.
Physical knots tied on polymer chains display contour lengths $l_k$ that scale with the length of the chain $N$ as $l_k \sim N^{\theta}$ with $0<\theta <1$ in dilute, good solvent conditions and with $\theta \simeq 1$ under isotropic confinement~\cite{Tubiana2011prl,Marcone2005,Tubiana2013}. For a static TopoII, the topological relaxation time is bound to be comparable to the (curvilinear) diffusion time of the knotted portion $\tau_R \sim (N-l_k)^2/D_c \sim (N-l_k)^2 l_k/D_0 \sim N^{2+\theta}/D_0$, in the limit $N \to \infty$ and where $D_0$ is the microscopic diffusion of a bead. On the other hand, a jumping TopoII only needs a timescale that scales as the (inverse) probability of landing on a segment that belongs to the knotted portion, i.e.~$\tau_T \sim k_j^{-1} (N/l_k) \sim N^{1-\theta}$. Clearly, this is a much weaker scaling in polymer length $N$ with respect to the static case. Remarkably, the more delocalised the knot, the closer $\theta$ to unity (e.g. under strong isotropic confinement or in bad solvent $\theta \simeq 1$~\cite{Tubiana2011prl,Virnau2005}) and the starker the difference in timescales, $\tau_R/\tau_T \sim N^{2\theta+1}$.

\section{Discussion} 
Understanding the various mechanisms through which Topoisomerase proteins regulate DNA and genome topology remains an open challenge in biophysics. 
\dmi{Type II Topoisomerase (TopoII) is an abundant nuclear protein (around $10^6$ copies per cell~\cite{Padget2005}) and is often found to be enriched near specific elements (such as CTCF~\cite{Uuskula-Reimand2016,Canela2017a} and cohesin~\cite{Vian2018a}) in vivo. In spite of this, the effect of TopoII dynamic binding on the topological relaxation of in vivo and in vitro substrates has not been properly addressed so far. To tackle these outstanding questions, here we have studied some previously underappreciated yet realistic aspects which may affect the action of TopoII on DNA: its dynamics and biased binding towards regions of large curvature or local density of the substrate. We discover that dynamic binding, with realistic values of dissociation constant $k_D \simeq 50$ nM~\cite{Charvin2005a} (corresponding to unbinding times of the order of 0.1-1 seconds, in fair agreement with FRAP data~\cite{Kalfalah2011,Christensen2002}) yields enhanced sampling of topological spaces which in turn drives faster topological simplification.}

\dmi{Interestingly, we find that this enhanced topological relaxation is accelerated in the presence of binding biases to regions of larger local density. We can justify these biases using simple thermodynamic considerations. Indeed, since TopoII has two DNA binding sites (associated with the ``gate'' and ``transfer'' segment~\cite{Bates2005}) it would be energetically preferable to have them both bound to DNA at the same time. This is most easily achieved in regions where there is a larger density of DNA sites thus supporting the idea that TopoII may have a thermodynamic preference to locally enrich regions of larger DNA density. We note that this is a generic result, and any DNA binding protein with $n>1$ non-specific binding sites would follow the same behaviour, as also suggested by the bridging-induced attraction model~\cite{Brackley2013,Ryu2021}. Accordingly, we do not need to invoke any specific role for ATP in the search process. Instead, our results are generic and our model will be able to complement any other more detailed model dissecting the role of ATP in the strand-crossing process~\cite{Burnier2007,Yan1999}. } 
\dmi{In line with this, we note that our simulations -- irrespectively of whether they model static or dynamic TopoII -- eventually converge to values of knotting probability that are equal or greater than the one expected in equilibrium (see SI, Fig.~S8).} 

Perhaps the most important and striking finding of this work is that introducing dynamic binding renders the topological relaxation process independent on the substrate length. To the best of our knowledge this aspect has never been tested experimentally and could be realised with time-resolved gel electrophoresis measurements of DNA topology in vitro such as the ones performed in Ref.~\cite{Stuchinskaya2009}. 

\dmi{We further highlight that the recent findings of Valdes et al~\cite{Valdes2018} suggest that in vivo chromatin attains a plateau in the knotting probability around 8-10 kb. If this behaviour were mirrored by the topological simplification rate, then our results would imply that TopoII would have residence times in vitro smaller or equal than $\tau_R \simeq 0.1$ seconds (obtained using $R_g \simeq 0.29 \mu$m and $D \simeq 1\mu$m$^2/s$ measured in vitro in free solution~\cite{Robertson2006}). In vivo, it is expected the effective viscosity of the solution to be higher, e.g. $\eta_n \simeq 100-1000$ cP in the nucleus~\cite{Rosa2008,Caragine2018}, thus yielding slower dynamics, e.g. $D \simeq 0.01\mu$m$^2/s$ and in turn longer maximum unbinding times $\tau_R \simeq 10$ seconds in line with FRAP data~\cite{Kalfalah2011,Christensen2002}. }

Interestingly, our results may also be used to determine the residency time of TopoII on a plasmid DNA substrate at a bulk level as follows: by considering different populations of DNA knots with different lengths, we expect that for DNA molecules whose $\tau_R(N) < k_{\rm off}^{-1}$ the topological relaxation rate should dependent on the substrate length. On the other hand, when $\tau_R(N) \geq k_{\rm off}^{-1}$ then the topological relaxation should become independent on the length. By coupling kinetic (time course) measures of topological relaxation~\cite{Stuchinskaya2009} with fluorescent single-molecule tracking~\cite{Robertson2006} it may thus be possible to quantify the residency time of TopoII on DNA under different conditions, e.g. salt, crowding etc., at a bulk scale.

\dmi{Finally, we note that in presence of crowding~\cite{Jeon2016} or bounded domains~\cite{Mattos2012}, the mean passage time (for instance mean unknotting time in the present problem) may be an insufficient measure of the underlying process; further, target search in the genome has notoriously broad arrival times~\cite{Bauer2015}. Interestingly, we find that the width of the distributions of first unknotting times strongly depend on both the binding kinetics and the search strategy (see SI Fig.~S9) with the maximum density search strategy being the one with narrower first unknotting times distributions. It would be interesting to experimentally test this prediction too in the future.} 

In conclusion, we argue that due to our surprising findings, our work will become a necessary complement to previous models building towards a comprehensive framework of TopoII action on DNA. Beyond bringing us closer to a full appreciation of the many facets involving TopoII-mediated topological simplification, we hope our simulations will stimulate targeted experiments on this fascinating long-standing problem.

\section{ACKNOWLEDGEMENTS}
This project has received funding from the European Research Council (ERC) under the European Union's Horizon 2020 research and innovation programme (grant agreement No 947918, TAP). DM also acknowledges support of the Royal Society via a University Research Fellowship. LT acknowledges support from MIUR, Rita Levi Montalcini Grant, 2016. Source codes are available at \href{https://git.ecdf.ed.ac.uk/ygutier2/topo2-lammps-implementation.git}{https://git.ecdf.ed.ac.uk/ygutier2/topo2-lammps-implementation.git}.

\bibliographystyle{nar}
\bibliography{biblio_fin}

\end{document}